\documentstyle[aps,epsf,rotate,preprint,tighten]{revtex}

\begin{document}

\draft

\title{Scaling behavior in economics: I. Empirical results \\ 
for company growth}

\author{Lu\'{\i}s~A.~Nunes Amaral,$^{1,2}$ Sergey~V. Buldyrev,$^1$
Shlomo Havlin,$^{1,3}$\\ Heiko Leschhorn,$^{1}$\protect\footnote{Present
Address: Theor. Physik III, Heinrich-Heine-Univ., D-40225 D\"usseldorf,
Germany.}  Philipp Maass,$^{1}$\protect\footnote{Present Address:
Fakult\"at f\"ur Physik, Universit\"at Konstanz, D-78434 Konstanz,
Germany.}  Michael~A.~Salinger,$^4$ H.~Eugene Stanley,$^1$\\ and
Michael~H.~R.~Stanley$^{1}$\protect\footnote{Present Address: Department
of Physics, MIT, Cambridge, MA 02139.}}

\address{$^1$Center of Polymer Studies and Department of Physics,\\
                Boston University, Boston, MA 02215, USA\\
         $^2$Institut f\"ur Festk\"orperforschung, Forschungszentrum
                J\"ulich, D-52425 J\"ulich, Germany\\
         $^3$Department of Physics, Bar-Ilan University, Ramat Gan,
                Israel\\
         $^4$School of Management, Boston University, Boston, 
                MA 02215, USA
}

\date{last revised: October 31, 1996; printed: \today}

\maketitle

\begin{abstract}

  We address the question of the growth of firm size. To this end, we
  analyze the Compustat data base comprising all publicly-traded United
  States manufacturing firms within the years 1974--1993.  We find that
  the distribution of firm sizes remains stable for the 20 years we
  study, i.e., the mean value and standard deviation remain
  approximately constant.  We study the distribution of sizes of the
  ``new'' companies in each year and find it to be well approximated by
  a log-normal. We find (i) the distribution of the logarithm of the
  growth rates, for a fixed growth period of one year, and for companies
  with approximately the same size $S$ displays an exponential form, and
  (ii) the fluctuations in the growth rates --- measured by the width of
  this distribution $\sigma_1$ --- scale as a power law with $S$,
  $\sigma_1\sim S^{-\beta}$.  We find that the exponent $\beta$ takes
  the same value, within the error bars, for several measures of the
  size of a company.  In particular, we obtain: $\beta=0.20\pm0.03$ for
  sales, $\beta=0.18\pm0.03$ for number of employees,
  $\beta=0.18\pm0.03$ for assets, $\beta=0.18\pm0.03$ for cost of goods
  sold, and $\beta=0.20\pm0.03$ for property, plant, \& equipment.

\end{abstract}


\section{Introduction}

Statistical physics has undergone many changes in emphasis during the
last few decades.  The seminal works of the '60s and '70s on critical
phenomena \cite{Fisher,Wilson} provided physicists with a new set of
tools to study nature \cite{Wilson,deGennes}.  Fields such as
biophysics, medicine, geomorphology, geology, evolution, ecology or
meteorology are now common areas of application of statistical physics.

In particular, several statistical physics research groups have turned
their attention to problems in economics \cite{Bak,Stanley2} and finance
\cite{Solomon,Bouchaud,stock,Ghashghaie,Levy,Bak1,Potters,Takayasu92x,
Hirabayashi93x}.
On the other hand, the concepts of statistical physics (e.g.,
self-organization) have started to penetrate the study of economics
\cite{Krugman}.  In this article, we extend the study of
Ref.~\cite{Stanley2} on the growth rate of manufacturing companies. One
motivation for the present study is the considerable recent interest in
economics in developing a richer theory of the firm
\cite{Gibrat,Coase,Hart,Simon,Baumol,Hymer,Cyert,Jensen,Ijiri,Lucas,
Jovanovic,Nelson,GolanTh,Varian,Holmstrom,Williamson,Milgrom,Radner,
Golan,Shapiro}.  In standard microeconomic theory, a firm is viewed as a
production function for transforming inputs such as labor, capital, and
materials into output \cite{Hart,Jensen,Varian}.  When dynamics are
incorporated into the model, the link between production in one period
and production in another arises because of investment in durable,
physical capital and because of technological change (which in turn can
arise from investments in research and development).  Recent work on
firm dynamics emphasizes the effect of how firms learn over time about
their efficiency relative to competitors \cite{Cyert,Pakes1,Pakes2}. The
production dynamics captured in these models are not, however, the only
source of actual firm dynamics.  Most notably, the existing models do
not account for the time needed to assemble the organizational
infrastructure needed to support the scale of production that typifies
modern corporations.

We studied all United States (US) manufacturing publicly-traded firms
from 1974 to 1993.  The source of our data is Compustat which is a
database on all publicly-traded firms in the US. Compustat obtains this
information from reports that publicly traded companies must file with
the US Securities and Exchange Commission.  The database contains a
large amount of information on each company.  Among the items included
are ``sales,'' ``cost of goods sold,'' ``assets,'' ``number of
employees,'' and ``property, plant, \& equipment.''

Another item provided for each company is the Standard Industrial
Classification (SIC) code.  In principle, two companies in the same
primary SIC code are in the same market; that is, they compete with
each other.  In practice, defining markets is extremely difficult
\cite{Scherer}.  More important for our analysis, virtually all modern
firms sell in more than one market.  Companies that operate in
different markets do report some disaggregated data on the different
activities.  For example, while Philip Morris was originally a tobacco
producer, it is also a major seller of food products (since its
acquisition of General Foods) and of beer (since its acquisition of
Miller Beer).  Philip Morris does report its sales of tobacco
products, food products, and beer separately.  However, companies have
considerable discretion in how to report information on their
different activities, and differences in their choices make it
difficult to compare the data across companies.

In this paper, the only use we make of the primary SIC codes in
Compustat is to restrict our attention to manufacturing firms.
Specifically, we include in our sample all firms with a major SIC code
from 2000--3999.  We do not use the data from the individual business
segments of a firm, nor do we divide up the sample according to primary
SIC codes.  We should acknowledge that this choice is at odds with the
mainstream of economic analysis.  In economics, what is commonly called
the ``theory of the firm'' is actually a theory of a business unit.  To
build on the Philip Morris example, economists would likely not use a
single model to predict the behavior of Philip Morris.  At the very
least, they would use one model for the tobacco division, one for the
food division, and one for the beer division.  Indeed, given the
available data, they might construct a completely separate model of,
say, the sales of Maxwell House coffee.  Absent any effect of the output
of one of Philip Morris' products on either the demand for or costs of
its other products, the models of the different components of the firm
would be completely separate.  Because the standard model of the firm
applies to business units, it does not yield any prediction about the
distribution of the size of actual, multi-divisional firms or their
growth rates.

On the other hand, the approach we take in this study is part of a
distinguished tradition.  First, there is a large body of work by
Economics Nobel laureate H. Simon \cite{Ijiri} and various co-authors
that explored the stochastic properties of the dynamics of firm growth.
Also, in a widely cited article (that nonetheless has not had much
impact on mainstream economic analysis), R. Lucas, also a Nobel
laureate, suggests that the distribution of firm size depends on the
distribution of managerial ability in the economy rather than on the
factors that determine size in the conventional theory of the firm
\cite{Lucas}.

In summary, the objective of our study is to uncover empirical scaling
regularities about the growth of firms that could serve as a test of
models for the growth of firms.  We find: (i) the distribution of the
logarithm of the growth rates for firms with approximately the same
size displays an exponential form, and (ii) the fluctuations in the
growth rates --- measured by the width of this distribution --- scale
as a power law with firm size.

The paper is organized as follows: In Sect. II, we review the
economics literature on the growth of companies.  In Sects. III and
IV, we present our empirical results for publicly-traded US
manufacturing companies.  Finally, in Sect. V, we present concluding
remarks and discuss questions raised by our results.

\section{Background}

In 1931, the French economist Gibrat proposed a simple model to explain
the empirically observed size distribution of companies \cite{Gibrat}.
He made the following assumptions: (i) the growth rate $R$ of a company
is independent of its size (this assumption is usually referred to by
economists as the {\it law of proportionate effect\/}), (ii) the
successive growth rates of a company are uncorrelated in time, and (iii)
the companies do not interact.

In mathematical form, Gibrat's model is expressed by the stochastic
process:
\begin{equation}
S_{t+ \Delta t} = S_t (1 + \epsilon_t),
\label{e-gibrat}
\end{equation}
where $S_{t+ \Delta t}$ and $S_t$ are, respectively, the size of the
company at times $(t+ \Delta t)$ and $t$, and $\epsilon_t$ is an
uncorrelated random number with some bounded distribution and variance
much smaller than one (usually assumed to be Gaussian).  Hence $\log
S_t$ follows a simple random walk and, for sufficiently large time
intervals $u\gg\Delta t $, the growth rates 
\begin{equation}
R_u\equiv\frac{S_{t+u}}{S_t}
\end{equation}
are log-normally distributed.  If we assume that all companies are
born at approximately the same time and have approximately the same
initial size, then the distribution of company sizes is also
log-normal. This prediction from the Gibrat model is approximately
correct \cite{Stanley1,HO}.

There is, however, considerable evidence that contradicts Gibrat's
underlying assumptions. The most striking deviation is that the
fluctuations of the growth rate measured by the relative standard
deviation $\sigma_1(S)$ decline with an increase in firm size. This was
first observed by Singh and Whittington \cite{Singh} and confirmed by
others \cite{Stanley2,Evans,Hall,Dunne,Davis1,Davis2}.  The negative
relationship between growth fluctuations and size is not surprising
because large firms are likely to be more diversified. Singh and
Whittington state that the decline of the standard deviation with size
is not as rapid as if the firms consisted of independently operating
subsidiary divisions.  The latter would imply that the relative standard
deviation decays as $\sigma_1(S)\sim S^{-1/2}$ \cite{Singh}.  This
confirms the common-sense view that the performance of different parts
of a firm are related to each other.

The situation for the mean growth rate is less clear. Singh and
Whittington \cite{Singh} consider the assets of firms and observe that
the mean growth rate increases slightly with size. However, the work of
Evans \cite{Evans} and Hall \cite{Hall}, using the number of employees
to define the company's size, suggests that the mean growth rate
declines slightly with size. Dunne et al. \cite{Dunne} emphasize the
effect of the failure rate of firms and the effect of the ownership
status (single- or multi-unit firms) on the relation between size and
mean growth rate.  They conclude that the mean growth rate is always
negatively related with size for single-unit firms; but for multi-unit
firms, the growth rate increases modestly with size because the
reduction in their failure rates overwhelms a reduction in the growth of
nonfailing firms \cite{Dunne}.

Another testable implication of Gibrat's law is that the growth rate of
a firm is uncorrelated in time. However, the empirical results in the
literature are not conclusive. Singh and Whittington \cite{Singh}
observe positive first order correlations in the 1-year growth rate of a
company (persistence of growth) whereas Hall \cite{Hall} finds no such
correlations.  The possibility of negative correlations (regression
towards the mean) has also been suggested \cite{Leonard,Friedman}.

\section{Size distribution of publicly-traded companies}

In the following sections, we study the distribution of company sizes
and growth rates.  To do so, one problem that must be confronted is
the definition of firm size.  If all companies produced the same good
(steel, say), then we could use a physical measure of output, such as
tons.  We are, however, studying companies that produce different
goods for which there is no common physical measure of output.  An
obvious solution to the problem is to use the dollar value of output:
the sales.  A general alternative to measuring the size of output is
to measure input.  Again, since companies produce different goods,
they use different inputs.  However, virtually all companies have
employees.  As a result, some economists have used the number of
employees as a measure of firm size.  Three other possibilities
involve the dollar value of inputs, such as the ``cost of goods
sold,'' ``property, plant \& equipment,'' or ``assets.''  As we
discuss below, we obtain similar results for all of these measures.
We begin by describing the growth rate of sales.  To make the values
of sales in different years comparable, we adjust all values to 1987
dollars by the GNP price deflator.

\begin{figure}
\centerline{
\epsfysize=0.8\columnwidth{\rotate[r]{\epsfbox{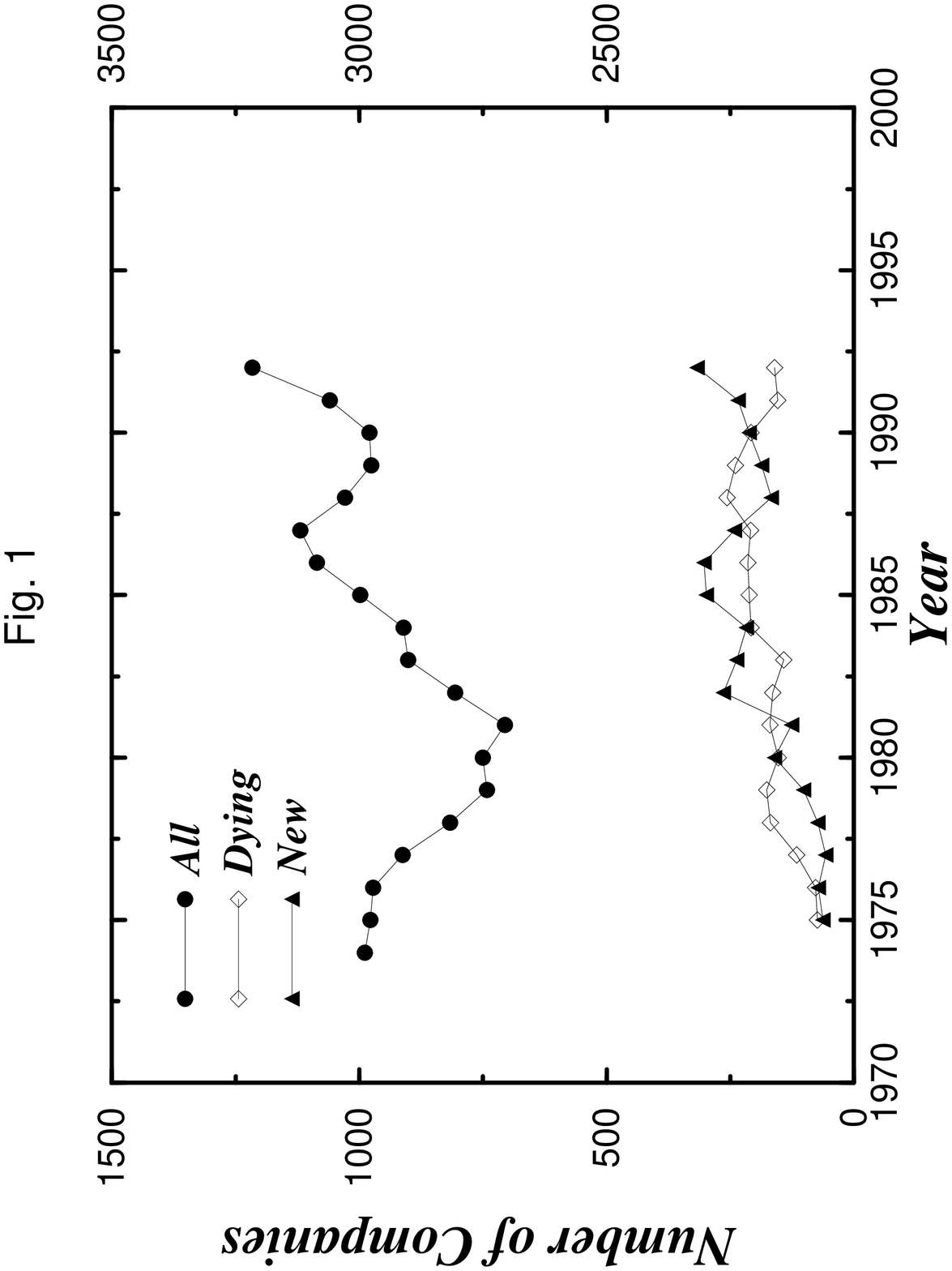}}}
}
\vspace*{1.0cm}
\caption{
        Number of publicly-traded manufacturing companies in the US
for the period 1974--1993 (right scale).  Also shown is the number of
companies entering the market and the number of companies leaving
(left scale).}
\label{f-number}
\end{figure}

Since the law of proportionate effect implies a multiplicative process
for the growth of companies, it is natural to study the logarithm of
sales.  We thus define
\begin{equation}
s_0 \equiv \ln S_0
\end{equation}
and the corresponding growth rate
\begin{equation}
r_1 \equiv \ln R_1 = \ln \frac{S_1}{S_0}, 
\end{equation}
where $S_0$ is the size of a company in a given year and $S_1$ its
size the following year.

Stanley et al. determined the size distribution of publicly-traded
manufacturing companies in the US \cite{Stanley1}.  They found that for
1993, the data fit to a good degree of approximation a log-normal
distribution. These results have been recently confirmed by Hart and
Oulton \cite{HO} for a sample of approximately $80000$ United Kingdom
companies. Here, we present a study of the distribution for a period of
20 years (from 1974 to 1993).

Figure~\ref{f-number} shows the total number of publicly-traded
manufacturing companies present in the database each year.  We also
plot the number of {\it new\/} companies and of ``{\it dying}''
companies (i.e., companies that leave the database because of merger,
change of name or bankruptcy).

Figure 2(a) shows the distribution of firm size in each year from
1974--1993. Particularly above the lower tails, the distributions lie
virtually on top of each other.
\begin{figure}
\narrowtext
\vspace*{.5cm}
\centerline{
\epsfysize=0.8\columnwidth{\rotate[r]{\epsfbox{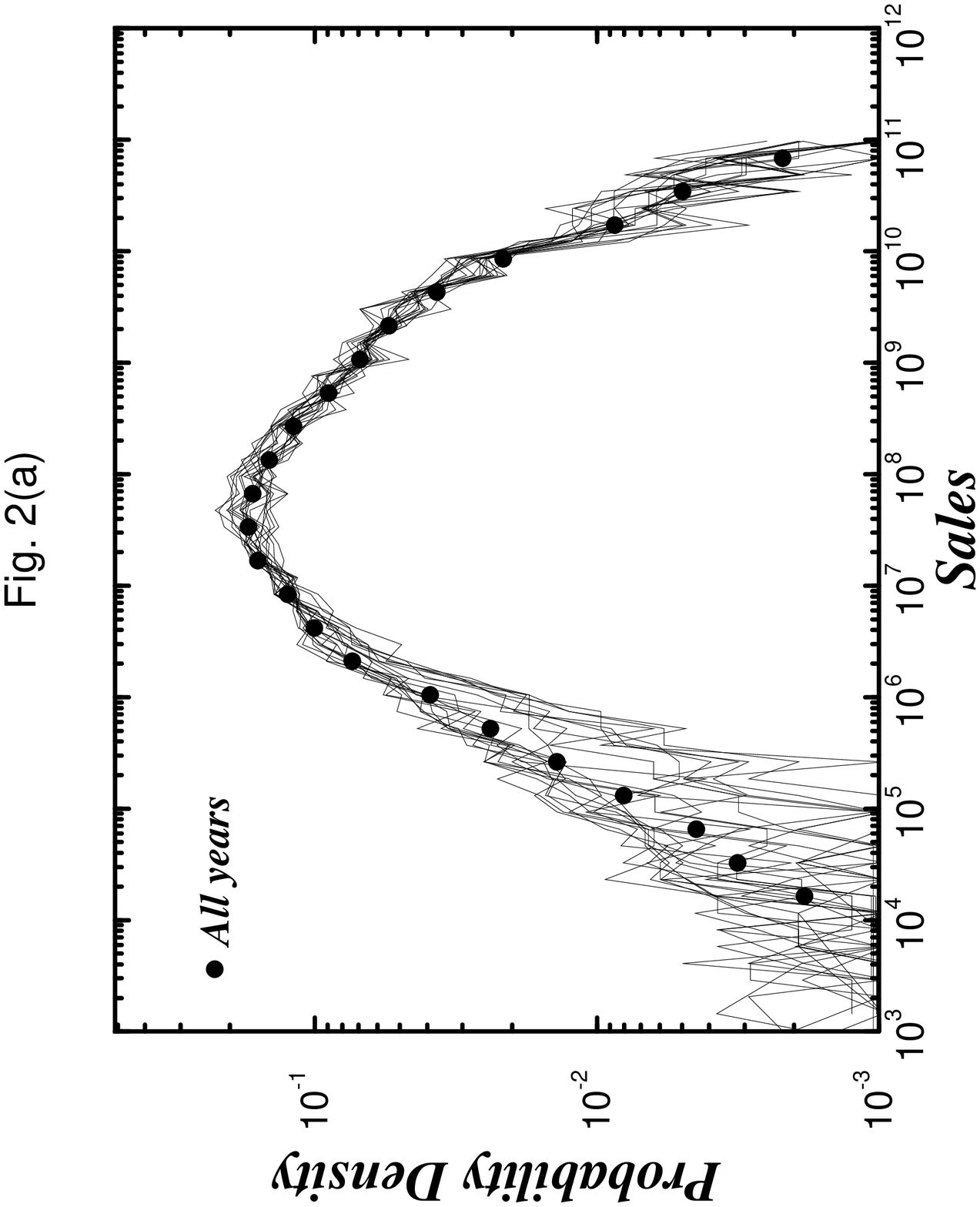}}}
}
\vspace*{1.5cm}
\centerline{
\epsfysize=0.8\columnwidth{\rotate[r]{\epsfbox{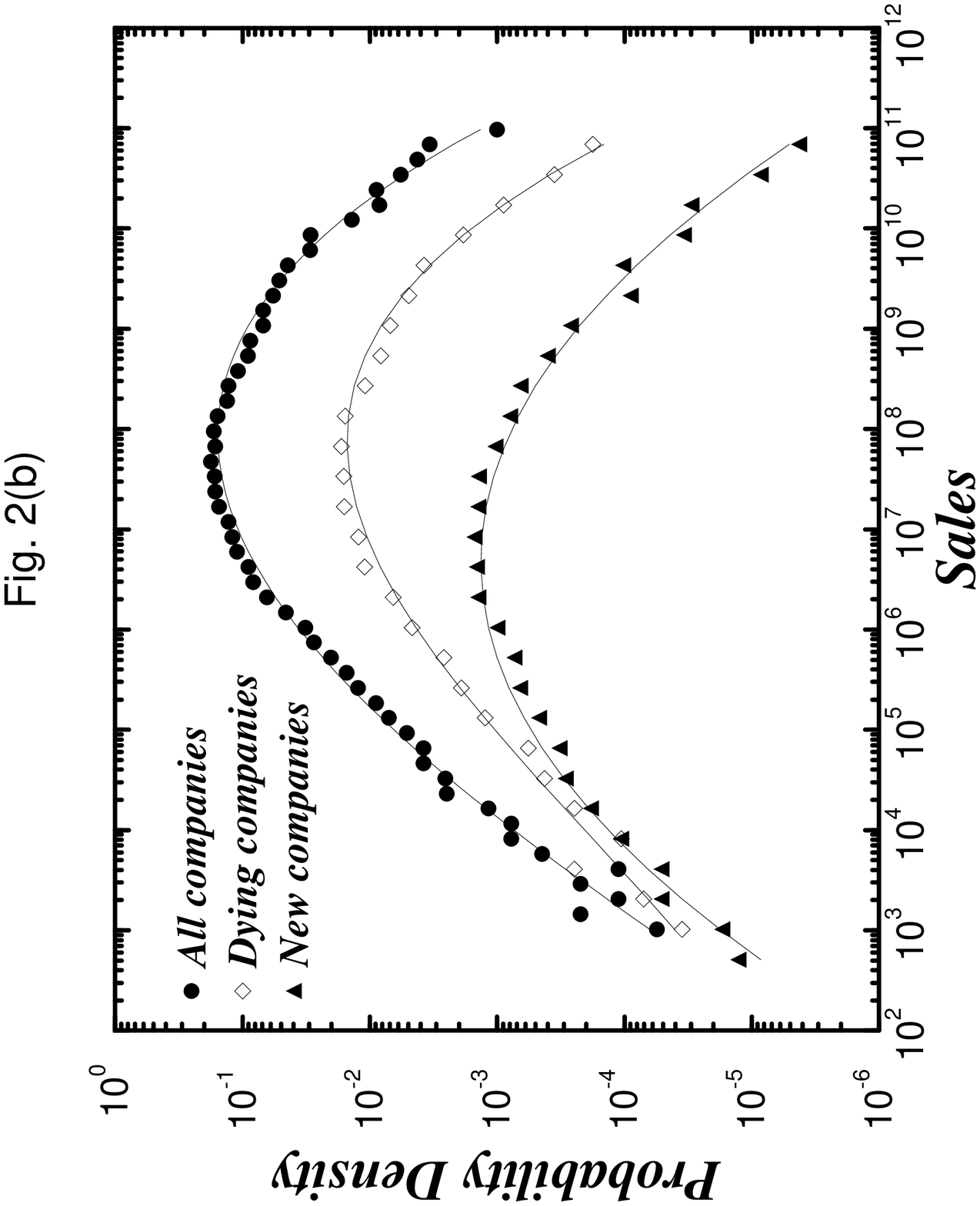}}}
}
\vfill
\centerline{
\epsfysize=0.8\columnwidth{\rotate[r]{\epsfbox{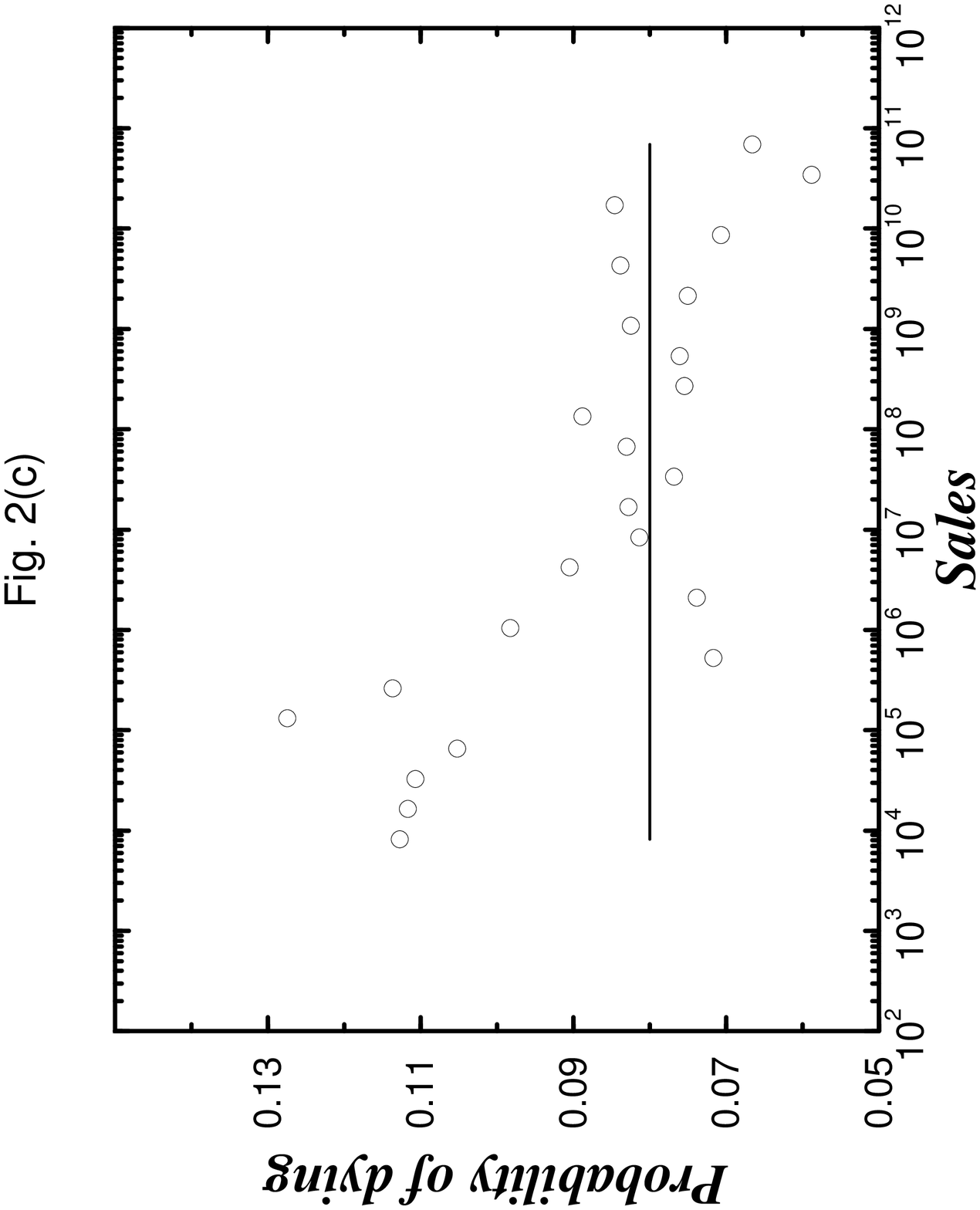}}}
}
\vspace*{1.0cm}
\caption{(a) Probability density of the logarithm of the sales for
        publicly-traded manufacturing companies (with standard
        industrial classification index of 2000-3999) in the US for {\it
        each\/} of the years in the 1974--1993 period.  All the values
        for sales were adjust to 1987 dollars by the GNP price deflator.
        Also shown (solid circles) is the average over the 20 years.  It
        is visually apparent that the distribution is approximately
        stable over the period.  (b) Probability density of the
        logarithm of sales for all the manufacturing companies, for the
        companies entering the market (shifted by a factor of $1/10$),
        and for the companies leaving the market (shifted by a factor of
        $1/100$), averaged over the 1974--1993 period.  The distribution
        of new companies can be described to first approximation by a
        log-normal while the other distributions are better fitted by
        the exponential of a third order polynomial.  Notice that the
        distributions of all companies and of dying companies are nearly
        identical.  This suggests a nearly constant dependence of the
        dying probability on size.  (c) Plot of the fraction of
        ``dying'' companies by size.  We define this probability as the
        ratio of dying companies of a given size over the total number
        of companies of that size. The horizontal straight line is a
        guide for the eye for companies with sales above $10^6$.}
\label{f-size}
\end{figure}
 Thus the distribution is stable over
this period. This is surprising because there is no existing theoretical
reason to expect that the size distribution of firms could remain stable 
value slightly smaller than the average of all companies.  One might
as the economy grows, as the composition of output changes, and as factors 
that economists would expect to affect firm size (like computer technology)
evolve.  It is also important because it contradicts the
predictions of the Gibrat model.  Equation (\ref{e-gibrat}) implies that
the distribution of sizes of companies should get broader with time.  In
fact, the variance of the distribution should increase linearly in time.
Thus, we must conclude that other factors, not included in Gibrat's
assumptions, must have important roles.

One obvious factor not captured by the Gibrat assumption is the entry of
new companies.  Figure 2(b) shows that the size distribution of new
publicly-traded companies is approximately a log-normal with an average
expect new companies to be much smaller on average than existing ones.
However, new companies can come about through the merger of two existing
companies, in which case the new company is bigger than either of the
pre-existing companies.  Another way that new companies come into
existence is that very large companies divest themselves of divisions
that are, by themselves, large businesses.  An example is AT\&T's recent
divestiture of its manufacturing division (Lucent) and its computer
division (NCR).

Another factor not included in Gibrat's assumptions is the ``dying''
of companies.  As shown in Fig.~\ref{f-size}b, this distribution is
quite similar to the distribution for all companies.  Thus, it
suggests that the probability for a company to leave the market,
whether by merger, change of name, or bankruptcy, is nearly
independent of size, Fig.~\ref{f-size}c.

When analyzing the data, it is important to consider the high level of
the noise in the tails.  In building a histogram from the data, the
most straightforward method is to use equally spaced bins.  However,
doing so creates noisy results in the tails because of the
small number of data points in these regions.  One way to solve this
problem, especially if some knowledge of the shape of the distribution
exists, is to take bins chosen with such lengths that all of them
receive approximately the same number of data points.  In fact, we
used equally spaced bins on a {\it logarithmic scale}, i.e., all firms
with sales values falling into an interval between $8^k$ and $8^{k+1}$
with $k$ an integer belong to one bin.

\section{The distribution of growth rates}

\begin{figure}
\narrowtext
\centerline{
\epsfysize=0.8\columnwidth{\rotate[r]{\epsfbox{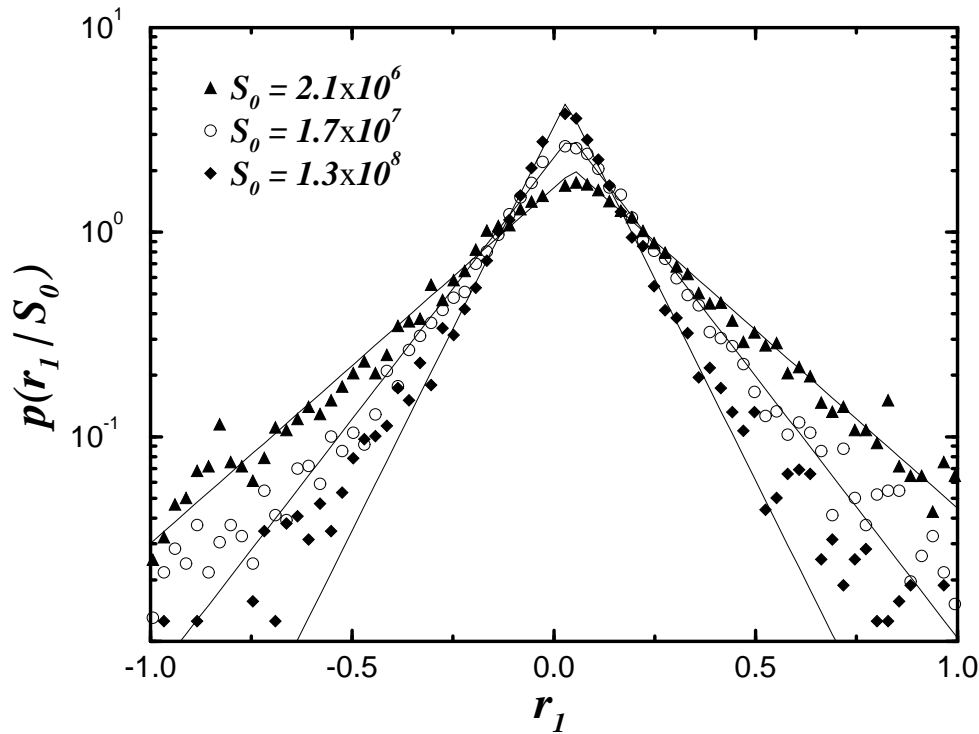}}}
}
\vspace*{1.0cm}
\caption{ Probability density $p(r_1|s_0)$ of the growth rate
        $r\equiv\ln(S_{1}/S_{0})$ for all publicly-traded US
        manufacturing firms in the 1994 Compustat database with Standard
        Industrial Classification index of 2000--3999. The
        distribution represents all annual growth rates observed in
        the 19-year period 1974--1993.  We show the data for three
        different bins of initial sales (with sizes increasing by
        powers of 8): $8^{7}<S_0<8^{8}$, $8^{8}<S_0<8^{9}$, and
        $8^{9}<S_0<8^{10}$.  Within each sales bin, each firm has a
        different value of $R$, so the abscissa value is obtained by
        binning these $R$ values.  The solid lines are exponential
        fits to the empirical data close to the peak.  We can see that
        the wings are somewhat ``fatter'' than what is predicted by an
        exponential dependence.}
\label{f-distribution}
\end{figure}

The distribution $p(r_1|s_0)$ of the growth rates from 1974 to 1993 is
shown in Fig.~\ref{f-distribution} for three different values of the
initial sales.  Remarkably, these curves can be approximated by 
a simple "tent-shaped"
form.  Hence the distribution is not Gaussian --- as expected from the
Gibrat approach \cite{Gibrat} --- but rather is exponential
\cite{Stanley2},
\begin{equation}
p(r_1|s_0)={1\over \sqrt{2}\sigma_1(s_0)} \exp \left(
-{\sqrt{2}\,|r_1-\bar r_1(s_0)|\over\sigma_1(s_0)}
\right).
\label{e-distribution}
\end{equation}
The straight lines shown in Fig.~\ref{f-distribution} are calculated
from the average growth rate $\bar r_1(s_0)$ and the standard deviation
$\sigma_1(s_0)$ obtained by fitting the data to
Eq.~(\ref{e-distribution}). The tails of the distribution in 
Fig.~\ref{f-distribution} are somewhat fatter than Eq.~(\ref{e-distribution})
predicts. This deviation is the opposite of what one would find if the 
distribution were Gaussian. We find that the data for {\it each\/}
annual interval from 1974--1993 also fit well to
Eq.~(\ref{e-distribution}), with only small variations in the parameters
$\bar r_1(s_0)$ and $\sigma_1(s_0)$.

\subsection{Mean growth rate}

\begin{figure}
\vspace*{.5cm}
\narrowtext
\centerline{
\epsfysize=0.8\columnwidth{\rotate[r]{\epsfbox{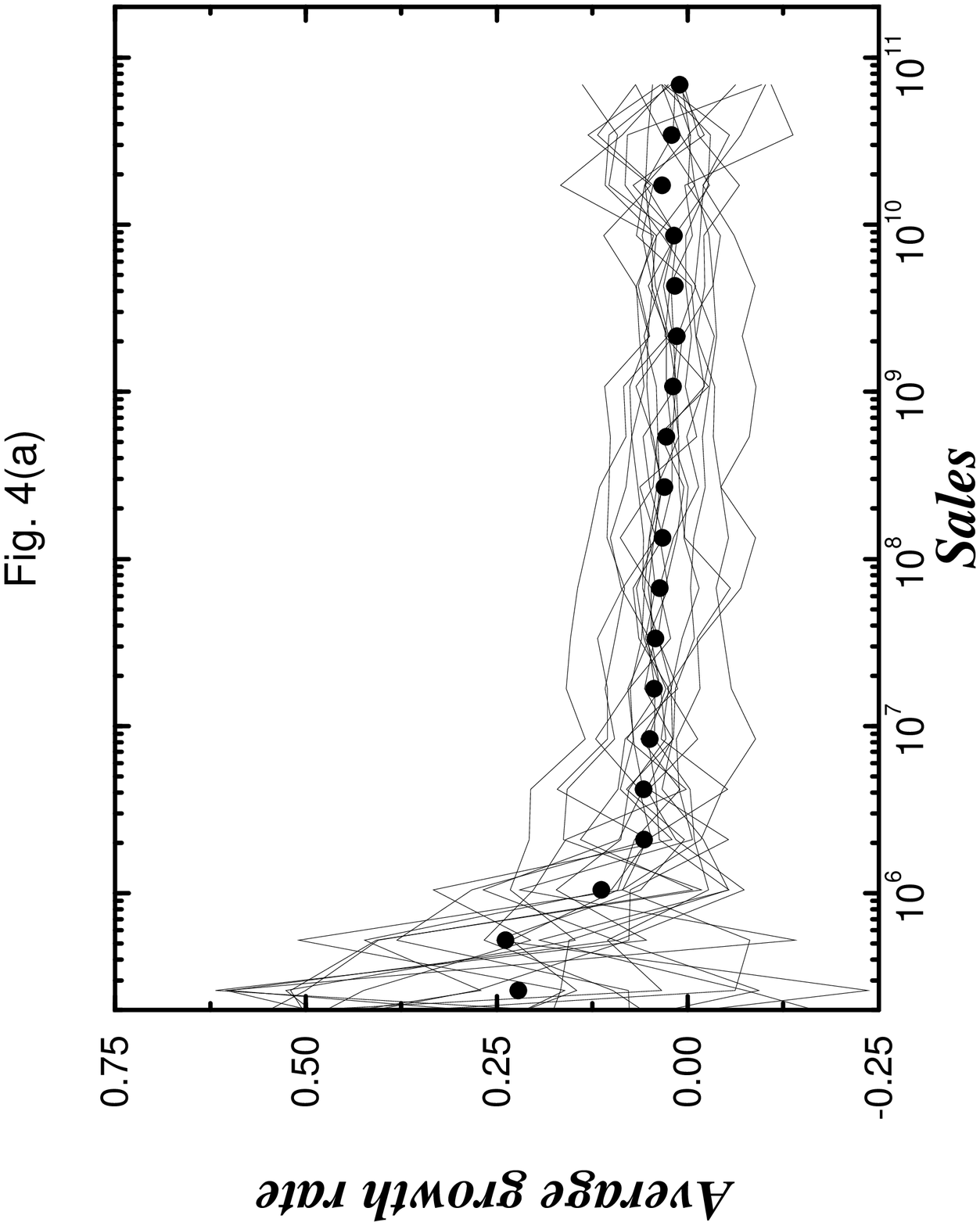}}}
}
\vspace*{1.5cm}
\centerline{
\epsfysize=0.8\columnwidth{\rotate[r]{\epsfbox{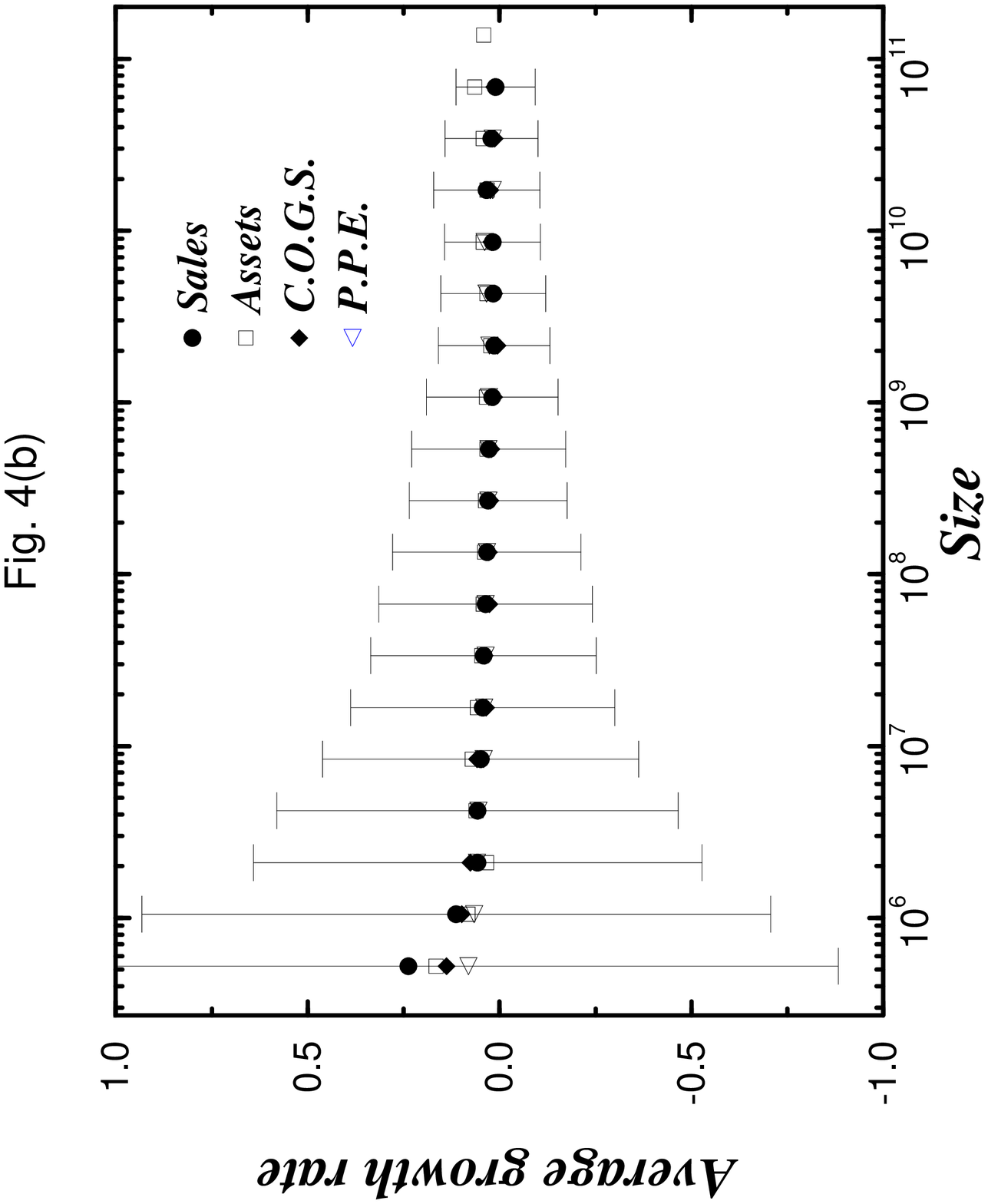}}}
}
\vfill
\centerline{
\epsfysize=0.8\columnwidth{\rotate[r]{\epsfbox{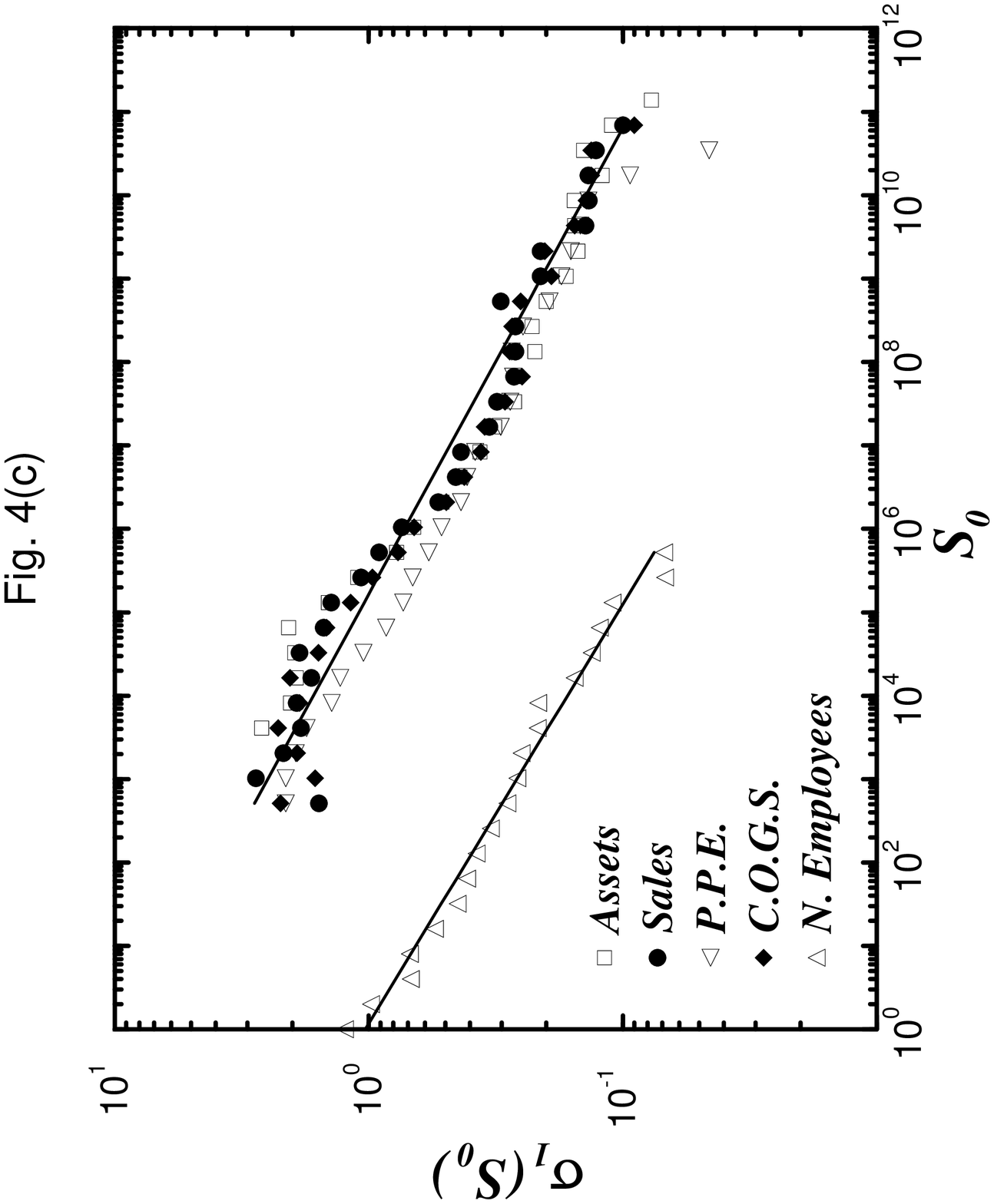}}}
}
\vspace*{1.0cm}
\caption{ (a) Mean 1-year growth rate $\bar r_1(s_0)$ for several
        years.  It is visually apparent that the data are quite noisy,
        and that there is no significant dependence on $S_0$ (at most
        a logarithmic dependence with a very small coefficient).  Also
        displayed is the mean growth rate for the 18-year period in
        Compustat.  (b) Average for the 19 years of $\bar r_1(s_0)$
        for several size definitions: sales, assets, cost of goods
        sold and plant property and equipment.  Error bars
        corresponding to one standard deviation are shown for sales
        --- values for the other quantities are nearly identical.
        Again, no significant dependence on $S_0$ is found.  Although
        it seems likely that the slightly positive value of $\bar
        r(s_0)$ is a real effect, we cannot rule out the possibility
        of a bias of the data towards successful companies.  (c)
        Standard deviation of the 1-year growth rates for different
        definitions of the size of a company as a function of the
        initial values.  Least squares power law fits were made for all
        quantities leading to the estimates of $\beta$: $0.18\pm 0.03$
        for ``assets,'' $0.20\pm 0.03$ for ``sales,'' $0.18\pm0.03$
        for ``number of employees,'' $0.18\pm0.03$ for ``cost of goods
        sold,'' and $0.20\pm0.03$ for ``plant, property \&
        equipment.''  The straight lines are guides for the eye and
        have slopes $0.19$}
\label{f-growth}
\end{figure}

Economists typically have studied the relationship between mean growth
rate and firm size by running a regression of growth rates on firm size
sometimes with other control variables included.  Rather than using
regression analysis, we undertake a graphical analysis of the mean
growth rate.  Figure~\ref{f-growth}(a) displays $\bar r(s_0)$ as a
function of initial size $s_0$ for several years.  Although the data are
quite noisy, they suggest that there is {\it no\/} significant
dependence of the mean growth rate on $s_0$.  Least squares fits of the
individual curves to a form $\bar r(s_0) \sim s_0$ lead to estimates
of the proportionality constant which are very small in magnitude
($<10^{-2}$), and whose sign can be positive or negative depending on
the year.  Our analysis suggests that if a dependence exists, it is {\it
very\/} weak for any range of sizes where other factors, such as a bias
of the sample towards successful companies, could be disregarded.

The analysis for the average of the nineteen 1-year periods, which is
displayed in Fig.~\ref{f-growth}(b), confirms this observation.
Furthermore, the figure suggests that the results do not change when
we consider other definitions of the size of a company.

\subsection{Standard deviation of the growth rate}

Next, we study the dependence of $\sigma_1(s_0)$ on $s_0$.
As is apparent from Figs.~\ref{f-distribution}-\ref{f-growth}, the
width of the distribution of growth rates decreases with increasing
$s_0$.  We find that $\sigma_1(s_0)$ is well approximated
for 8 orders of magnitude (from sales of less than $10^3$ dollars up
to sales of more than $10^{11}$ dollars) by the law \cite{Stanley2}
\begin{equation}
\sigma_1(s_0) \sim \exp(-\beta s_0),
\label{e-sigma}
\end{equation}
where $\beta = 0.20 \pm 0.03$.  Equation (\ref{e-sigma}) implies the
scaling law
\begin{equation}
\sigma_1(S_0) \sim {S_0}^{-\beta}.
\label{e-sigma1}
\end{equation}
Figure~\ref{f-growth}(c) displays $\sigma_1$ vs. $S_0$, and we can see
that Eq.~(\ref{e-sigma1}) is indeed verified by the data.

\subsection{Other Measures of Size}

In order to test further the robustness of our findings, we perform a
parallel analysis for the number of employees.  We find that the
analogs of $p(r_1|s_0)$ and $\sigma_1(s_0)$ behave
similarly.  For example, Fig.~\ref{f-growth}(c) shows the standard
deviation of the number of employees, and we see that the data are
linear over roughly 5 orders of magnitude, from firms with less than
$10$ employees to firms with almost $10^6$ employees. The slope $\beta
= 0.18\pm 0.03$ is the same, within the error bars, as found for the
sales.

As shown in Fig.~\ref{f-growth}(c), we find that
Eqs.~(\ref{e-distribution}) and (\ref{e-sigma1}) approximately describe
three additional indicators of a company's size, (i) assets (with
exponent $\beta = 0.18 \pm 0.03$) (ii) cost of goods sold ($\beta =
0.18\pm 0.03$) and (iii) property, plant \& equipment ($\beta = 0.20\pm
0.03$).

\section{Discussion}

What is remarkable about Eqs.~(\ref{e-distribution}) and
(\ref{e-sigma1}) is that they approximate the growth rates of a diverse set
of firms. They range not only in their size but also in what they
manufacture. The conventional economic theory of the firm is based on
production technology, which varies from product to product.
Conventional theory does not suggest that the processes governing the
growth rate of car companies should be the same as those governing,
e.g., pharmaceutical or paper firms.
\begin{figure}
\narrowtext
\centerline{
\epsfysize=0.8\columnwidth{\rotate[r]{\epsfbox{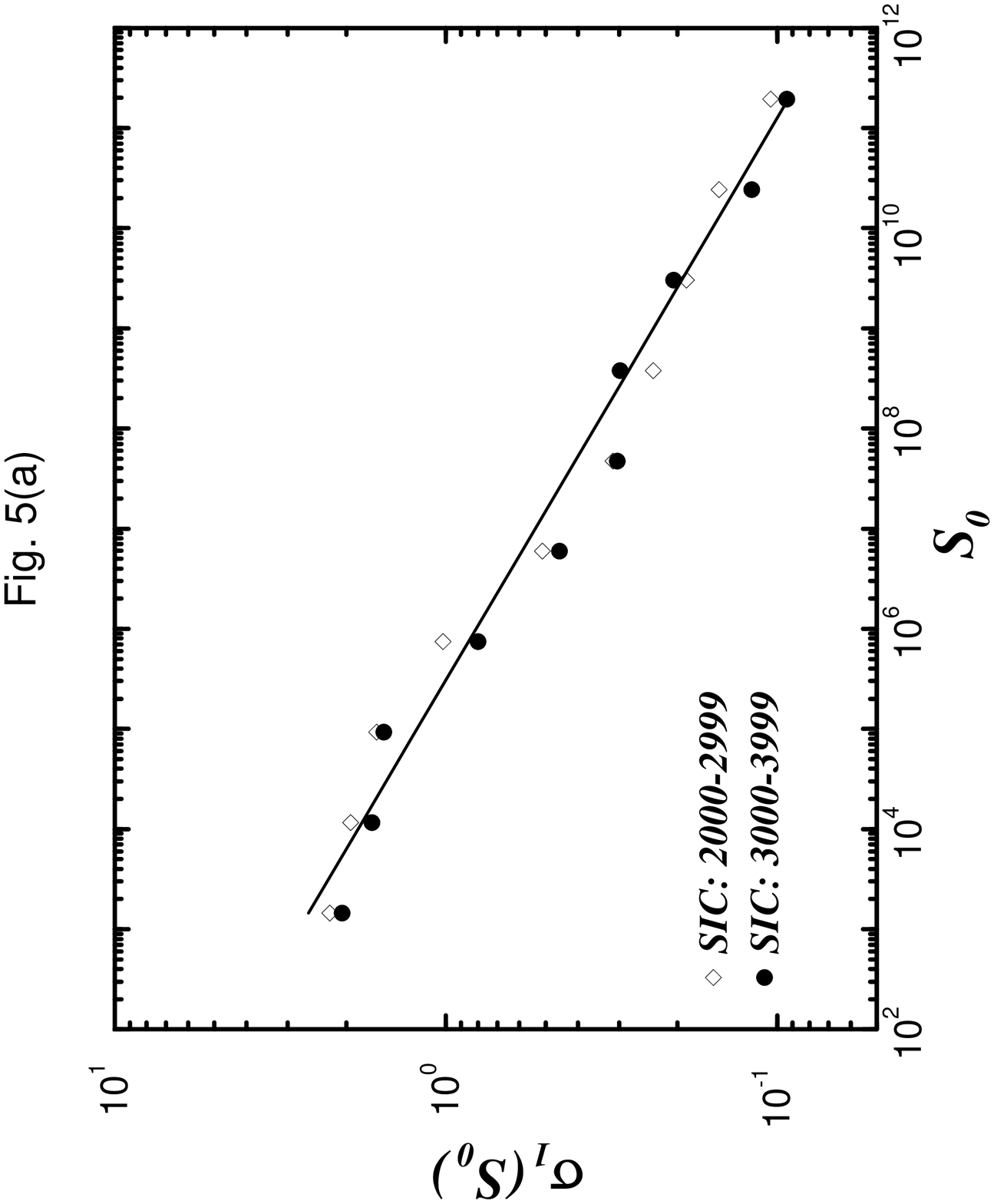}}}
}
\vfill
\vspace*{1.5cm}
\centerline{
\epsfysize=0.8\columnwidth{\rotate[r]{\epsfbox{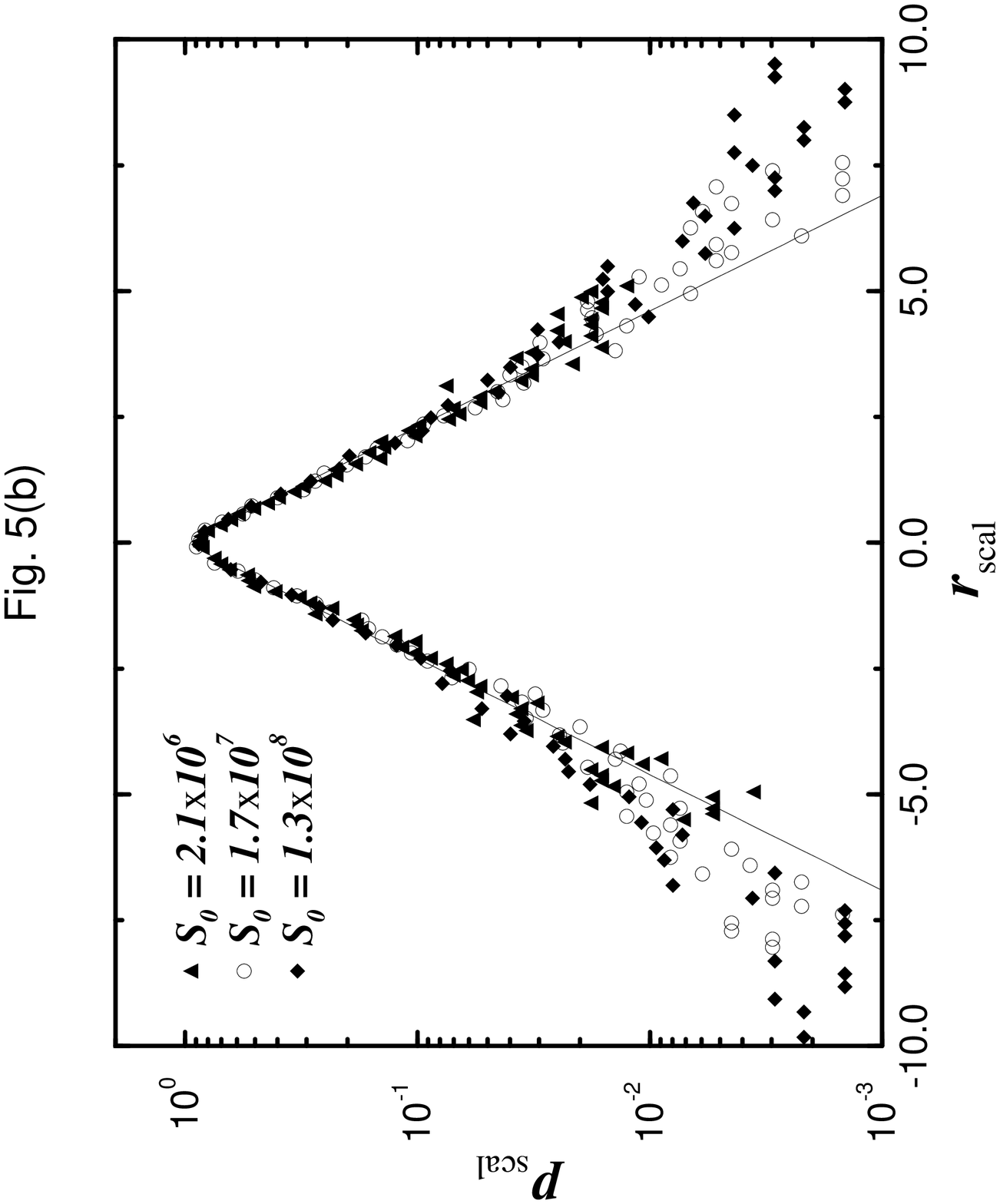}}}
}
\vspace*{1.0cm}
\caption{
        (a) Dependence of $\sigma_1$ on $S_0$ for two subsets of the
data corresponding to different values of the SIC codes.  In principle,
companies in different subsets operate in different markets.  The figure
suggests that our results are universal across markets.  (b) Scaled
probability density $p_{\mbox{\scriptsize scal}} \equiv \protect\sqrt{2}
\sigma_1(s_0) p(r_1|s_0)$ as a function of the scaled growth rate
$r_{\mbox{\scriptsize scal}} \equiv \protect\sqrt{2} [r_1-\bar r_1(s_0)]
/ \sigma_1(s_0)$.  The values were rescaled using the measured values of
$\bar r_1({\rm s}_0)$ and $\sigma_1(s_0)$.  All the data collapse upon
the universal curve $p_{\mbox{\scriptsize scal}} = \exp
(-|r_{\mbox{\scriptsize scal}}|)$ as predicted by
Eqs.~(\protect\ref{e-distribution}) and (\protect\ref{e-sigma}).}
\label{f-collapse}
\end{figure}
  Indeed, our findings are
reminiscent of the concept of universality found in statistical physics,
where different systems can be characterized by the same fundamental
laws, independent of ``microscopic'' details.  Thus, we can pose the
question of the universality of our results: Is the measured value of
the exponent $\beta$ due to some averaging over the different
industries, or is it due to a universal behavior valid across all
industries? As a ``robustness check,'' we split the entire sample into
two distinct intervals of SIC codes.  It is visually apparent in
Fig.~5(a) that the same behavior holds for the different samples of 
industries.

In statistical physics, scaling phenomena of the sort that we have
uncovered in the sales and employee distribution functions are
sometimes represented graphically by plotting a suitably ``scaled''
dependent variable as a function of a suitably ``scaled'' independent
variable.  If scaling holds, then the data for a wide range of
parameter values are said to ``collapse'' upon a {\it single} curve.
To test the present data for such data collapse, we plot in
Fig.~\ref{f-collapse}(b) the scaled probability density
$p_{\mbox{\scriptsize scal}} \equiv \sqrt{2}\sigma(s_0)p(r_1|{\rm
s}_0)$ as a function of the scaled growth rates of both sales and
employees $r_{\mbox{\scriptsize scal}}\equiv \sqrt{2}[r_1-\bar
r_1(s_0)]/\sigma(s_0)$.  The data collapse relatively well upon the single
curve $p_{\rm scal}=\exp(-|r_{\mbox{\scriptsize scal}}|)$.  Our
results for (i) cost of goods sold, (ii) assets, and (iii) property,
plant \& equipment are equally consistent with such scaling.
The high degree of similarity in the behavior of sales, the number of
employees, and of the other measures of size that we studied points to
the existence of large correlations among those quantities, as one would
expect.

In summary, we study publicly-traded US manufacturing companies from
1974 to 1993.  We find that the distribution of the logarithms of the
growth rate decays exponentially.  Furthermore, we observe that the
standard deviation of the distribution of growth rates scales as a power
law with the size $S$ of the company.  Our results support the
possibility that the scaling laws used to describe complex but inanimate
systems comprised of many interacting particles (as occurs in many
physical systems) may be usefully extended to describe complex but
animate systems comprised of many interacting subsystems (as occurs in
economics).  Furthermore, the kind of scaling laws found in this study
can be viewed as empirical evidence supporting some hypothesis regarding
the self-organization of the economy \cite{Krugman}.

\acknowledgments

  We thank R. N. Mantegna for important help in the early stages of
this work, and JNICT (L.A.), DFG (H.L. and P.M.), and NSF for
financial support.


\begin{thebibliography}{99}

\bibitem{Fisher} M.~E. Fisher, Rep. Progr. Phys. {\bf 30}, 615--731
(1967); M.~E. Fisher, {\it Critical Phenomena\/} [Proc. 1970 Enrico
Fermi Int. School of Physics, Course No. 51, Varenna, Italy], edited by
M. S. Green (Academic Press for Ital. Phys. Soc., New York, 1971),
pp.~1-99.

\bibitem{Wilson}
K.~G. Wilson, {\it The 1982 Nobel Lectures\/} (World Scientific Press,
Singapore, 1982).

\bibitem{deGennes} P.-G. de Gennes, {\it The 1991 Nobel Lectures\/}
(World Scientific Press, Singapore, 1991); {\it Scaling Concepts in
Polymer Physics\/} (Cornell University Press, Ithaca, 1979).

\bibitem{Mandelbrot}
B.~B. Mandelbrot, {\it The Fractal Geometry of Nature\/}
(W.H. Freeman, New York, 1983).

\bibitem{Bak}
P. Bak, K, Chen, J.~A. Scheinkman and M. Woodford, Richerche
Economichi {\bf47}, 3 (1993); J.~A. Scheinkman and J. Woodford,
American Economic Review {\bf 84} 417 (1994).

\bibitem{Stanley2}
M.~H.~R. Stanley, L.~A.~N. Amaral, S.~V. Buldyrev, S. Havlin,
H. Leschhorn, P. Maass, M.~A. Salinger, and H.~E. Stanley, Nature 
{\bf 379}, 804 (1996).

\bibitem{Solomon}
M. Levy, H. Levy, and S. Solomon, Economics Letters {\bf 45}, 103
(1994).

\bibitem{Bouchaud}
J.-P. Bouchaud and D. Sornette, J. Phys. I (France) {\bf 4}, 863
(1994); D. Sornette, A. Johansen, and J.-P. Bouchaud, J. Phys. I
(France) {\bf 6}, 167 (1996).

\bibitem{stock}
R.~N. Mantegna  and H.~E. Stanley, Nature {\bf 376}, 46 (1995).

\bibitem{Ghashghaie} S. Ghashghaie, W. Breymann, J. Peinke, P. Talkner,
and Y. Dodge, Nature {\bf 381}, 767 (1996); see also R.~N. Mantegna and
H.~E. Stanley, Nature {\bf 383}, 587 (1996), and A. Arneodo et
al. (preprint).

\bibitem{Levy}
M. Levy and S. Solomon, ``Power laws are logarithmic Boltzmann laws'' 
(preprint, 1996). 

\bibitem{Bak1}
P. Bak, M. Paczuski, and M. Shubik, ``Price variations in a stock
market with many agents'' (preprint, 1996).

\bibitem{Potters}
M. Potters, R. Cont, J.-P.  Bouchaud, ``Financial markets as adaptative 
systems'' (preprint, 1996).

\bibitem{Takayasu92x}
H. Takayayasu, H. Miura, T. Hirabayashi, and K. Hamada, Physica A {\bf
  184}, 127--134 (1992).

\bibitem{Hirabayashi93x}
T. Hirabayashi, H. Takayayasu, H. Miura, and K. Hamada, Fractals {\bf
  1}, 29--40 (1993).

\bibitem{Krugman}
P.~R. Krugman, {\it The Self-Organizing Economy\/} (Blackwell
Publishers, Cambridge, 1996).

\bibitem{Gibrat}
R. Gibrat, {\it Les In\'egalit\'es Economiques\/} (Sirey, Paris,
1931).

\bibitem{Coase}
R.~H. Coase, Economica {\bf 4}, 386 (1937).

\bibitem{Hart}
P.~E. Hart and S.~J. Prais, J. Royal Statistical. Society, Series
{\bf A 119}, 150 (1956).

\bibitem{Simon}
H.~A. Simon and C.~P. Bonini, American Economical Review {\bf
48}, 607 (1958).

\bibitem{Baumol}
W. Baumol, {\it Business Behavior, Value, and Growth\/} (MacMillan,
New York, 1959).

\bibitem{Hymer}
S. Hymer and P. Pashigian, J. Political Economics {\bf 52},
556 (1962).

\bibitem{Cyert}
R. Cyert and J. March. {\it A Behavioral Theory of the Firm\/}
(Prentice-Hall, Englewood Cliffs, New Jersey, 1963).

\bibitem{Jensen}
M.~C. Jensen and W.~H. Meckling, Journal Financial Economics. {\bf 3},
305 (1976).

\bibitem{Ijiri}
Y. Ijiri and H.~A. Simon, {\it Skew Distributions and the Sizes
of Business Firms\/} (North Holland, Amsterdam, 1977).

\bibitem{Lucas}
R. Lucas, Bell J. Economics {\bf 9}, 508 (1978).

\bibitem{Jovanovic}
B. Jovanovic,  Econometrica {\bf 50}, 649 (1982).

\bibitem{Nelson}
R.~R. Nelson and S.~G. Winter, {\it An Evolutionary Theory of Technical
Change\/} (Harvard University Press, Cambridge, Massachusetts, 1982).

\bibitem{GolanTh}
A. Golan, {\it A Discrete Stochastic Model of
Economic Production and a Model of Fluctuations in Production 
-- Theory and Empirical Evidence\/} (Ph.D. Thesis, University of
California, Berkeley, 1988).

\bibitem{Varian}
H.~R. Varian, {\it Microeconomics Analysis\/} (Norton, New York,
1978).

\bibitem{Holmstrom}
B.~R. Holmstrom and J. Tirole, in {\it Handbook of Industrial
Organization\/} Vol. 1, eds. R. Schmalensee and R. Willig, 61
(North Holland, Amsterdam, 1989).

\bibitem{Williamson}
O.~E. Williamson, in {\it Handbook of Industrial Organization\/} 
Vol. 1, eds. R. Schmalensee and R. Willig, 135 (North Holland,
Amsterdam, 1989).

\bibitem{Milgrom}
P. Milgrom and J. Roberts, {\it Economics, Organization, and Management}
(Prentice-Hall, Englewood Cliffs, New Jersey, 1992).

\bibitem{Radner}
R. Radner, Econometrica {\bf 61}, 1109 (1993).

\bibitem{Golan}
A. Golan, Advances in Econometrics {\bf 10}, 1 (1994).

\bibitem{Shapiro}
D. Shapiro, R.~D. Bollman, and P. Ehrensaft, American J.
Agricultural Economics {\bf 69}, 477 (1987).

\bibitem{Pakes1}
A. Pakes and P. McGuire, Rand Journal {\bf 25}, 555 (1994).

\bibitem{Pakes2}
R.~E. Ericson and A. Pakes, Review of Economic Studies {\bf 62}, 
53 (1995).

\bibitem{Scherer}
F.~M. Scherer and D.~R. Ross, {\it Industrial Market Structure and
Economic Performance\/} (Houghton Mifflin, Boston, 1990).

\bibitem{Stanley1}
M.~H.~R. Stanley, S.~V. Buldyrev, R. Mantegna, S. Havlin,
M.~A. Salinger, and H.~E. Stanley, Economics Letters {\bf 49}, 453
(1995).

\bibitem{HO}
P. Hart and N. Oulton, The Economic Journal (in press).

\bibitem{Singh}
A. Singh and G. Whittington, Review Economical Studies {\bf 42}, 15
(1975).

\bibitem{Evans}
D.~S. Evans, J. Political Economics {\bf 95}, 657 (1987).

\bibitem{Hall}
B.~H. Hall, J. Industrial Economics {\bf 35}, 583 (1987).

\bibitem{Dunne}
T. Dunne, and M. Roberts, and L. Samuelson, Quaterly J. Economics
{\bf 104}, 671 (1989).

\bibitem{Davis1}
S.~J. Davis and J. Haltiwanger, Quaterly Journal Economics {\bf107},
819 (1992).

\bibitem{Davis2}
S.~J. Davis, J. Haltiwanger, and S. Schuh, {\it Job Creation and
Destruction\/} (MIT Press, Cambridge, Massachusetts, 1996).

\bibitem{Leonard}
J. Leonard, National Bureau of Economic Research, working
paper no. 1951, (1986).

\bibitem{Friedman}
M. Friedman, J. Economical Literature {\bf 30}, 2129 (1992).

\end{thebibliography}
\end{document}